\documentclass[amsmath,amssymb,amsbsy,reprint,prb,preprintnumbers,showpacs,superscriptaddress]{revtex4-2}
\usepackage{graphicx,color}
\usepackage{dcolumn}
\usepackage{bm}
\usepackage{braket}
\usepackage{mathtools}
\usepackage{ulem}
\usepackage[breaklinks,colorlinks=true,linkcolor=blue,urlcolor=blue,citecolor=blue]{hyperref}
\usepackage{times}
\usepackage{physics}
\usepackage{latexsym}
\usepackage{amsmath, amssymb}
\usepackage{mathtools}
\usepackage{multirow}

\newcommand{\be}{\begin{eqnarray}}
\newcommand{\ee}{\end{eqnarray}}

\renewcommand{\theequation}{\arabic{equation}}

\begin{document}
\title{
Strong-to-weak spontaneous symmetry breaking and average symmetry protected topological order in the doubled Hilbert space}

\date{\today}
\author{Yoshihito Kuno} 
\thanks{These authors equally contributed to this work}
\affiliation{Graduate School of Engineering science, Akita University, Akita 010-8502, Japan}
\author{Takahiro Orito}
\thanks{These authors equally contributed to this work}
\affiliation{Institute for Solid State Physics, The University of Tokyo, Kashiwa, Chiba, 277-8581, Japan}
\author{Ikuo Ichinose} 
\thanks{A professor emeritus}
\affiliation{Department of Applied Physics, Nagoya Institute of Technology, Nagoya, 466-8555, Japan}


\begin{abstract} 
Discovering and categorizing quantum orders in mixed many-body systems are currently one of the most important problems. 
Target model in this study is an extended version of the cluster model in one dimension with $Z_2\otimes Z_2$ symmetry, and we investigate effects of decoherence
applied to the ground state of the model, focusing on the symmetry aspect.
By using a scheme that we propose, a strong symmetry protected topological (SPT) mixed state and double average SPT (ASPT) state are constructed
through the pure gapless SPT order and the domain-wall duality.
Among them, the double ASPT is categorized by coexisting orders, i.e., a strong-to-weak spontaneous symmetry breaking and ASPT defined by the remaining weak and strong symmetries.
We make use of the doubled Hilbert space formalism for the construction scheme.
We numerically demonstrate the emergence of the two mixed SPT states and find that a phase transition occurs between them tuned by the strength of decoherence. 
Finally, we discuss the coexistence of SPT and SWSSB in the double SPT state from the view point of symmetrically invertible property, and
comment on the classification of ASPT proposed recently.
Suitable multiple-decoherence channel applied to SPT states gives a broad possibility to induce rich ASPTs, possessing non-trivial internal entanglement properties
from the view of doubled Hilbert space formalism.
\end{abstract}


\maketitle
\section{Introduction}
Environmental effects such as noise and decoherence \cite{gardiner2000} in open quantum systems are unavoidable in realistic situations. 
In quantum computers and quantum memories, interactions with the environment lead to undesirable effects, altering quantum states in an unfavorable manner~\cite{preskill2018,dennis2002,wang2003,ohno2004}.

However, even in the presence of environmental effects, noisy intermediate-scale quantum devices \cite{ebadi2021,bluvstein2024} are expected to demonstrate computational capabilities that surpass classical systems \cite{preskill2018}. 
Surprisingly, noise and decoherence can also induce rich physical phenomena, leading to non-trivial mixed quantum states that have no direct counterparts in isolated quantum systems. 
In particular, when noise and decoherence are applied to pure quantum states, they can give rise to exotic mixed quantum states, which may play a crucial role in quantum devices.

Recently, discovering non-trivial mixed state quantum matter in quantum many-body system is on-going issue. 
It is an open question to classify the mixed states in condensed matter physics as well as quantum information theory. 
In previous work, some interesting non-trivial mixed many-body state has been already investigated. As one of concrete examples, a topologically-ordered pure state \cite{Wen_text,wen2004} tends to change to a mixed state with another type of topological order \cite{bao2023,Wang_2025,sohal2024,zhang2024,sang2024,Chen2024_v2,kuno_2025_v1}.
Also, the behavior of symmetry protected topological (SPT) states under decoherence has been studied and some mixed state SPT or its pure SPT robustness to decoherence were clarified \cite{Lee2025,Guo-and-Ashida2024,min2024}. 
In these mixed states, two types of symmetries—strong and weak—can be defined \cite{groot2022,Buca2012} (see Appendix A). The interplay between these symmetries leads to a rich classification of mixed-state SPTs and the possibility of realizing intrinsically new SPT phases with no counterpart in the pure-state framework \cite{ma2023,ma2024,Ma2024_double}. In particular, SPT states protected by an averaged form of symmetry (weak symmetry) \cite{groot2022,Buca2012} for the density matrix have been investigated \cite{ma2023,ma2024,Ma2024_double}. In this context, the group cohomology classification method for pure-state SPTs \cite{Chen2013,Chen2014} has been extended \cite{Ma2024_double}, leading to the concept of averaged SPTs (ASPTs).

Then for mixed states, the notion of spontaneous symmetry breaking (SSB) is also extended, and there are various SSB patterns due to the presence of the notion of strong and weak symmetries. 
Some concrete examples have already been reported as several types of SSB in various systems, called strong symmetry SSB, weak symmetry SSB,
strong to weak SSB (SWSSB), etc.~\cite{lee2023,lessa2024,sala2024,KOI2024,sohal2024,liu2024_SSSB,guo2024,shah2024,weinstein2024,Ando2024,chen2024,stephen2024,Wang_2025,Orito2025}.

However, several open questions remain: How do spontaneous symmetry breakings (SSBs) and averaged symmetry-protected topological states (ASPTs) in mixed states relate to each other? Is there a possible coexistence of these SSBs and ASPTs in mixed states?

Quite recently, further studies addressing these questions have been posted in \cite{Guo2024_1,Guo2024_2}. Within the Lindblad formalism, it has been shown that certain steady states can exhibit the coexistence of strong-weak spontaneous symmetry breaking (SWSSB) and ASPT, forming what is referred to as the "double ASPT" phase. 
This phase emerges under appropriately chosen Lindblad jump operators and Hamiltonian dynamics. 
The construction of such steady states draws an analogy to the approach used for generating gapless SPT states in the pure-state framework, as introduced in \cite{Scaffidi2017}. 
In addition, the gSPT is demonstrated in a circuit dynamics \cite{Yu2025}.

In this work, we further predict the co-existence of SWSSB and ASPT mixed state by considering a construction scheme applying quantum channels inducing the SWSSB, not Lindblad dynamics. 
This construction protocol is based on the one of the gapless SPT (gSPT) in the pure-state framework \cite{Scaffidi2017}. 

Combined with this approach, 
we make use of Choi isomorphism technique and the doubled Hilbert space formalism~\cite{Choi1975,JAMIOLKOWSKI1972},
which is especially valid to consider symmetry properties of certain specific mixed states ~\cite{lee2023,Lee2025}.
Then, we deduce non-trivial decohered phases from a SPT pure system via decoherence, including
strong SPT phase and double ASPT phase and possibly gapless ASPT phase, similar to the steady states proposed in \cite{Guo2024_1,Guo2024_2}.
Each decohered state emerges as varying the decoherence strength. 

We further numerically validate our predictions by employing a filtering scheme for matrix product states (MPS) based on the doubled Hilbert space formalism \cite{Orito2025}. In this scheme, decoherence is applied to the mixed-state vector as a local filtering operation, a technique previously used in the analysis of pure states under perturbations, particularly for MPS in frustration-free models~\cite{Haegeman2015,Zhu2019}.

By using this scheme, the predicted phases are verified and we investigate the SWSSB, average SPT orders by observing some physical quantities, recently proposed in \cite{lessa2024,Ma2024_double,Guo2024_2}. 
In particular, through studying the entanglement properties for the doubled system, we clearly demonstrate an explicit phase transition behavior between strong-SPT and double ASPT, 
which can be predicted in the domain wall (DW) duality picture~\cite{Orito2025}. 
This transition is interesting in that the on-site decoherence to the one-dimensional system, in general, tends not to exhibit mixed state phase transition by tuning 
the strength of a decoherence parameter \cite{Lee2025}. 
Additionally, we explore possible edge mode degeneracy by analyzing the entanglement spectrum within the doubled Hilbert space framework. This analysis provides further insights into the bulk-edge correspondence in ASPTs, as predicted in the doubled Hilbert space formalism \cite{Ma2024_double}.

The rest of this paper is organized as follows. 
In Sec.~II, we review the construction scheme and protocol of the gSPT. 
In Sec.~III, based on this scheme, we propose its extensive application to a mixed state, where our target system under decoherence is introduced. 
We explain the doubled Hilbert space formalism and show the interpretation of the decoherence in this formalism. 
We expect the emergence of non-trivial mixed SPTs. 
In Sec.~IV, we show the systematic numerical demonstrations by using the MPS and the filtering to the MPS for various decoherence parameters. 
Here, we verify the properties of the non-trivial mixed SPTs and a phase transition between them.
In Sec.~V, we briefly discuss the coexistence of SPT and SWSSB in the double SPT state from the view point of symmetrically invertible property, and
comment on the classification of ASPT proposed recently.
Section VI is devoted for summary and conclusion.

\section{Construction of non-trivial mixed SPT}
We start to follow the construction scheme of the gapless SPT phase given in \cite{Scaffidi2017}. 
To this end, we shall consider the spin chain, Hamiltonian of which is composed of two kinds of spins, i.e., $\tau$-spin and $\sigma$-spin,
\begin{eqnarray}
H_0=\sum^{L-1}_{j=0}[-\tau^x_{j+\frac{1}{2}}-\sigma^x_{j}-J_{zz}\sigma^z_{j}\sigma^z_{j+1}],
\label{H0}
\end{eqnarray}
where $\tau$-spin resides on link $\{j+\frac{1}{2}\}$ and $\sigma$-spin on site $\{j\}$, the schematic image of which is shown in the upper panel of Fig.~\ref{Fig_system}. 
The system possesses $Z^\sigma_2\otimes Z^\tau_2$ symmetry, the generators of which are global spin-flip operators $\prod^{L-1}_{j=0}\sigma^x_{j}$ and 
$\prod^{L-1}_{j=0}\tau^x_{j+{1\over 2}}$, respectively. 
$Z^\sigma_2$ SSB is induced by increasing the parameter $J_{zz}$. 
For $J_{zz}=1$, the $\sigma$-spin enters a critical state of the Ising model (described by Ising conformal field theory). 
On the other hand, the $\tau$-spin is in a trivial product state completely decoupled from the $\sigma$-spin sector. 

As explained in Refs.~\cite{Chen2014,Scaffidi2017}, 
the system described by $H_0$ is converted into an SPT system protected by the $Z^\sigma_2\otimes Z^\tau_2$ symmetry. 
This is achieved by unitary transformation constituted by the product of the controlled-Z gates, namely domain-wall (DW) duality transformation \cite{Raussendorf2001,Tantivasadakarn2022,Scaffidi2017}, 
\begin{eqnarray}
U_{DW}\equiv\prod^{L-1}_{j=0}CZ_{j-1,j-\frac{1}{2}}CZ_{j,j-\frac{1}{2}},
\end{eqnarray}
where $CZ_{j-1,j-\frac{1}{2}}$ represents CZ gate for nearest-neighbor site and links, $(j-1,j+\frac{1}{2})$. 
Then, the Hamiltonian $H_0$ is converted as,
\begin{eqnarray}
&&H^{DW}_{0}\equiv U_{DW}H_0U^\dagger_{DW}\nonumber\\
&&=\sum_{j}[-\sigma^z_{j}\tau^x_{j+\frac{1}{2}}\sigma^z_{j+1}-\tau^z_{j-\frac{1}{2}}\sigma^x_{j}\tau^z_{j+\frac{1}{2}}-J_{zz}\sigma^z_{j}\sigma^z_{j+1}].
\end{eqnarray}
In the dual model, $\sigma$-spin and $\tau$-spin couple with each other and possibly generate entanglement. 
In other words, the unitary $U_{DW}$ changes properties of states inducing short-range entanglement. 
For $0<J_{zz}<1$, the ground state of $H^{DW}_{0}$ is a cluster SPT state protected by $Z^\sigma_2\otimes Z^\tau_2$ \cite{Raussendorf2001}, 
the state of which is obtained from the ground state of $H_0$ as 
$
|{\rm SPT_{cl}}\rangle=U_{DW}[|\tau^x\rangle|\sigma^{x}\rangle],
$
where $|\tau^x\rangle|\sigma^{x}\rangle$ is a product $x$ state for the $\tau$ and $\sigma$-spins.

Furthermore, by the DW duality transformation $U_{DW}$, it is expected that a gSPT and an SSB states appear as the ground state of $H^{DW}_{0}$. 
The gSPT state emerges through unitary $U_{DW}$ from the critical state of $H_0$ with $J_{zz}=1$, that is,
\begin{eqnarray}
|{\rm gSPT}\rangle=U_{DW}[|\tau^x\rangle|\sigma_{c}\rangle],
\end{eqnarray}
where the state $|\tau^x\rangle$ is simple $x$-product state of $\tau$-spins and $|\sigma_{c}\rangle$ is the Ising critical ground state of the $\sigma$-spin sector of $H_0$. 
The $|{\rm gSPT}\rangle$ is non-trivial in that it is gapless in the bulk, and if we consider an open system, 
the edge mode emerges and is robust to perturbations preserving the $Z^\sigma_2\otimes Z^\tau_2$ symmetry \cite{Scaffidi2017}. 

Then for $J_{zz}>1$, a specific type of SSB states appear as the ground state of $H^{DW}_{0}$ given by 
\begin{eqnarray}
|{\rm DW}_{\sigma SSB}\rangle=U_{DW}[|\tau^x\rangle|\sigma^{\pm}_{SSB}\rangle],
\end{eqnarray}
where the state $|\sigma^{\pm}_{SSB}\rangle$ is an SSB ground state of the $\sigma$-spin sector of $H_0$ with parity $\prod^{L-1}_{j=0}\sigma^x_{j}=\pm 1$. 
Although the $\tau$ and $\sigma$-spins are coupled within the $Z^\sigma_2$ SSB, the $\tau$-spin is still featureless (trivial) and, 
$Z^\tau_2$-trivial $\otimes Z^\sigma_2$ SSB state emerges \cite{Guo2024_1,Guo2024_2}.

Here, we note that the emergence of the states $|{\rm gSPT}\rangle$ and $|{\rm DW}_{\sigma SSB}\rangle$ comes from the Ising $\sigma^z\sigma^z$ interaction in
the Hamiltonian $H_0$. 
In general, to construct a non-trivial state through the DW duality transformation, some interaction parameters in the original Hamiltonian
are suitably chosen for an SSB state to appear in the subsystem of the original model before the DW duality transformation.

The above prescription gives insight into how to construct non-trivial mixed states with topological order as we show in the following section.

\section{Construction of non-trivial mixed SPT: Double ASPT}
In this section, we show that by applying the protocol explained in the previous section to a system under decoherence, mixed SPT states with some kind of SSB will be produced. 

For the pure state case in the previous section, we utilized the Ising interaction that induces the $Z^\sigma_2$ SSB to produce nontrivial states, in particular the SPTs. 
In the study on mixed SPT state, decoherence plays a similar role to the Ising interaction; 
we apply two types of decoherence to the $\sigma$-spin given by \cite{Nielsen2011} 
\begin{eqnarray}
\mathcal{E}_{\sigma^z\sigma^z}&=&\prod^{L-1}_{j=0}\mathcal{E}_{\sigma^z\sigma^z,j}\\
\mathcal{E}_{\sigma^z\sigma^z,j}[\rho]&=&\biggr[(1-p_{zz})\rho+p_{zz}\sigma^z_j\sigma^z_{j+1}\rho \sigma^z_{j+1}\sigma^z_{j}\biggl],\nonumber\\ 
\mathcal{E}_{\sigma^x}&=&\prod^{L-1}_{j=0}\mathcal{E}_{\sigma^x,j},\\
\mathcal{E}_{\sigma^x,j}[\rho]&=&\biggr[(1-p_{x})\rho+p_{x}\sigma^x_{j}\rho \sigma^x_{j}\biggl], \nonumber
\end{eqnarray}
where $\rho$ is a density matrix,  $p_{zz(x)}$ tunes the strength of decoherence and takes the value within $[0,1/2]$. 
For $p_{zz(x)}=1/2$, the decoherence channel is reduced to (non-selective) projective.

The order in application of local channels is irrelevant as long as we consider decoherence channels using Pauli operators 
such as $\mathcal{E}_{g_j}[\rho]=(1-p)\rho+g_j \rho g^\dagger_j$, where $g_j$ is an element of Pauli group with a finite length support.
Generally, two channels commute with each other, $\mathcal{E}_{g_j}\circ \mathcal{E}_{g_\ell}[\rho]=\mathcal{E}_{g_\ell}\circ \mathcal{E}_{g_j}[\rho]$ 
for either $[g_j,g_\ell]=0$ or $\{g_j,g_\ell\}=0$, and then, the operation of $\mathcal{E}_{\sigma^z\sigma^z,j}$ and $\mathcal{E}_{\sigma^x,j'}$ 
commute with each other for any $j$ and $j'$.

We consider the pure state density matrix $\rho_0\equiv |\psi_0\rangle\langle \psi_0|$, where the state $|\psi_0\rangle$ is the ground state of $H_0$ [Eq.(\ref{H0})]. 
Then, a decohered mixed state is given as,
\begin{eqnarray}
\rho_D\equiv \mathcal{E}_{\sigma^x}\circ\mathcal{E}_{\sigma^z\sigma^z}[\rho_0].
\label{rhoD}
\end{eqnarray}
Since the channels apply only $\sigma$ spins and do not have any effects on the $\tau$-spin, we can focus on only the $\sigma$-spin sector. 
In the $\sigma$-spin sector, the mixed state exhibits $Z^\sigma_{2}$ SWSSB for large $p_{zz}$ 
which has been clarified by the authors in \cite{Orito2025} with fixing $p_x=1/2-(1/2)(1-2p_{zz})^{1/J_{zz}}$, 
and the emergent SWSSB state still possesses 
the $Z^\sigma_{2}$ weak symmetry as well as strong $Z^\tau_2$ symmetry~\cite{ma2023,Ma2024_double}.
That is, a pair of decoherence operations $\mathcal{E}_{\sigma^z\sigma^z,j}$ and $\mathcal{E}_{\sigma^x,j'}$ surely induces the SWSSB analogous
to the action of the Ising interaction in the pure-state system of $H_0$. 
The parameter $p_{zz}$ plays a role of $J_{zz}$ in the pure-state system in the previous section. 
This remaining weak $Z^\sigma_2$ symmetry plays a key role in the emergence of an ASPT.

Similarly to the pure state case, for the mixed state framework, we will apply the DW duality transformation, and 
the transformation couples the $\tau$-spin with $\sigma$-spin sectors. 

\begin{figure}[t]
\begin{center} 
\includegraphics[width=8cm]{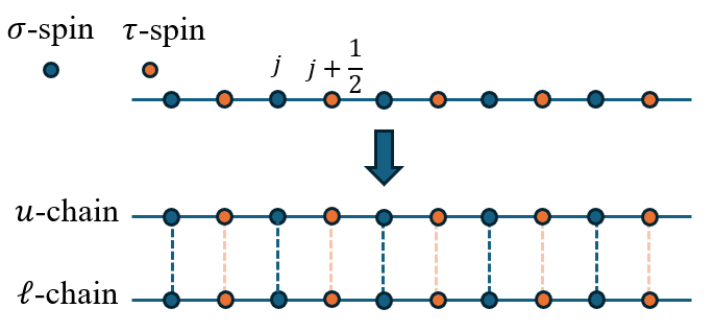}  
\end{center} 
\caption{Schematic image of the system: (Upper) One-dimensional $\sigma$ and $\tau$-spin system. 
(Lower) System in the doubled Hilbert space is constituted by the same upper($u$) and lower($\ell$) $\sigma$ and $\tau$-spin chains. 
The deocoherence operators induce the coupling between the upper and lower chains.}
\label{Fig_system}
\end{figure}

Before considering the DW duality transformation, as a useful theoretical framework for the treatment of mixed states, 
we introduce Choi isomorphism technique and the doubled Hilbert state formalism~\cite{Choi1975,JAMIOLKOWSKI1972}.
This formalism is numerically tractable in matrix product state simulation \cite{Orito2025}. We introduce the doubled Hilbert space, in which the target original Hilbert space $\mathcal{H}$ ($\tau$-spin plus $\sigma$-spin Hilbert spaces) is doubled as $\mathcal{H}_{u}\otimes \mathcal{H}_{\ell}$, where the subscripts $u$ and $\ell$ denote the upper (bra) and lower (ket) Hilbert spaces. The image is shown in the lower panel of Fig.~\ref{Fig_system}. 
Under this formalism, the pure state $\rho_0$ is described as \cite{Choi1975,JAMIOLKOWSKI1972}
\begin{eqnarray}
\rho_0\longrightarrow |\rho_0\rangle\rangle=|\psi^*_0\rangle|\psi_0\rangle. 
\end{eqnarray}
The state $|\rho_0\rangle\rangle$ is in the doubled Hilbert space $\mathcal{H}_u\otimes \mathcal{H}_{\ell}$.

In this formalism, a general decoherence channel $\mathcal{E}$ is mapped to a (linear) operator $\hat{\mathcal{E}}$ acting on a state vector $|\rho\rangle\rangle$ in 
$\mathcal{H}_{u}\otimes \mathcal{H}_{\ell}$ ~\cite{lee2023,Lee2025} and denoted as $\hat{\mathcal{E}}|\rho\rangle\rangle$.
Then, the quantum channels $\mathcal{E}_{\sigma^z\sigma^z}$ and $\mathcal{E}_{\sigma^x}$ become operators acting on $|\rho_0\rangle\rangle$ given explicitly by,
\begin{eqnarray}
\hat{\mathcal{E}}_{\sigma^z\sigma^z}(p_{zz})&=&\prod^{L-1}_{j=0}\biggr[(1-p_{zz})\hat{I}_{j,u}^* \otimes \hat{I}_{j,\ell}\nonumber\\ 
&&+p_{zz}\sigma^{z}_{j,u}\sigma^z_{j+1,u}\otimes \sigma^z_{j,\ell}\sigma^z_{j+1,\ell}\biggl]\nonumber\\ 
&=&\prod^{L-1}_{j=0}(1-2p_{zz})^{1/2}e^{\tau_{zz} \sigma^z_{j,u}\sigma^z_{j+1,u}\otimes \sigma^z_{j,\ell}\sigma^z_{j+1,\ell}},  
\nonumber
\end{eqnarray}
\begin{eqnarray}
\hat{\mathcal{E}}_{\sigma^x}(p_x)&=&\prod^{L-1}_{j=0}\biggr[(1-p_{x})\hat{I}_{j,u}^* \otimes \hat{I}_{j,\ell} +p_{x}\sigma^x_{j,u}\otimes \sigma^x_{j,\ell}\biggl]\nonumber\\
&=&\prod^{L-1}_{j=0}(1-2p_{x})^{1/2}e^{\tau_{x} \sigma^x_{j,u}\otimes \sigma^x_{j,\ell}},\nonumber
\end{eqnarray}
where, $\hat{I}_{j,u(\ell)}$ is an identity operator for site-$j$ vector space in $\mathcal{H}_{u(\ell)}$, $\sigma^z(\sigma^x)_{j,u(\ell)}$ is Pauli-$\sigma^z$($\sigma^x$) operator at site $j$ and $\tau_{zz(x)}=\tanh^{-1}[{p_{zz(x)}/(1-p_{zz(x)})}]$. 
Under this operation, the decohered state $\rho_D$ in Eq.~(\ref{rhoD}) is given by
\begin{eqnarray}
|\rho_D\rangle\rangle=\hat{\mathcal{E}}_{\sigma^x}(p_x)\hat{\mathcal{E}}_{\sigma^z\sigma^z}(p_{zz})|\rho_0\rangle\rangle.
\end{eqnarray}
We note that the channel operator $\hat{\mathcal{E}}$ is not unitary in general although the channel itself is a completely-positive 
trace-preserving map~\cite{Nielsen2011,lee2023,Lee2025}. 
Thus, the application of the channel operator generally changes the norm of the state vector.

In particular, $p_{zz}$-dependence of the decohered state of only $\sigma$-spin system in $|\rho_D\rangle\rangle$ has already been investigated numerically in \cite{Orito2025}. 
The $\sigma$-spin system exhibits the explicit $Z^\sigma_2$ SWSSB, which can be regarded as regime II in the quantum Ashkin teller model\cite{Ashkin1943,Kohmoto1981,Yamanaka1994}. 
Thus, the emergence of the SWSSB phase in the present system is expected.

We turn to apply the DW duality transformation for the mixed system in the doubled Hilbert space formalism in an extensive way. 

In applying the duality transformation $U_{DW}$, we use two step protocol; the first one is the DW transformation for $|\psi_0\rangle\rangle$,
\begin{eqnarray}
|\rho^{DW}_{0}\rangle\rangle \equiv U^*_{DW,u}\otimes U_{DW,\ell}|\rho_0\rangle\rangle.\nonumber
\end{eqnarray}
The state $|\rho^{DW}_{0}\rangle\rangle$ is the doubled state of the ground state of the Hamiltonian $H^{DW}_{0}$.
At the second step, we apply the duality to the quantum channels $\mathcal{E}_{\sigma^z\sigma^z}$ and $\mathcal{E}_{\sigma^x}$. 
Then, the channel $\mathcal{E}_{\sigma^z\sigma^z}$ is invariant but $\mathcal{E}_{\sigma^x}$ is transformed as
\begin{eqnarray}
&&(U^*_{DW,u}\otimes U_{DW,\ell})\hat{\mathcal{E}}_{\sigma^x}(U^*_{DW,u}\otimes U_{DW,\ell})^\dagger\nonumber\\
&&=\prod^{L-1}_{j=0}
\biggr[(1-p_x)\hat{I}_{j,u}^* \otimes \hat{I}_{j,\ell}\nonumber\\
&&+p_x\tau^z_{j-\frac{1}{2},u}\sigma^x_{j,u}\tau^z_{j+\frac{1}{2},u}
\otimes \tau^z_{j-\frac{1}{2},\ell}\sigma^x_{j,\ell}\tau^z_{j+\frac{1}{2},\ell}\biggl]\nonumber\\
&&\equiv \hat{\mathcal{E}}_{\tau^z\sigma^x\tau^z}.
\end{eqnarray} 
Then, we focus on the decohered state defined as 
\begin{eqnarray}
|\rho^{DW}_D\rangle\rangle&\equiv&\hat{\mathcal{E}}_{\tau^z\sigma^x\tau^z}(p_x)\hat{\mathcal{E}}_{\sigma^z\sigma^z}(p_{zz})|\rho^{DW}_0\rangle\rangle.
\end{eqnarray} 

The state $|\rho^{DW}_D\rangle\rangle$ can be regarded as the DW dual state of $|\rho_{D}\rangle\rangle$. 
The correspondence $|\rho_{D}\rangle\rangle \longleftrightarrow|\rho^{DW}_D\rangle\rangle$ is analogous to the relation between 
the ground state of $H_0$ and that of $H^{DW}_0$ in the pure state system including gSPT states \cite{Scaffidi2017}.

By the above protocol, we would like to construct non-trivial decohered states $|\rho^{DW}_D\rangle\rangle$;
\begin{enumerate}
\item {\bf Strong-symmetry cluster SPT mixed state}: For small $p_{zz}$ case corresponding to $0<J_{zz}<1$ in the pure state construction, the decohered state $|\rho^{DW}_D\rangle\rangle$ is similar to 
the pure state $|{\rm SPT}_{cl}\rangle$ in the sense that the Ising interaction and the decoherence effects are irrelevant. 
In the mixed state, the state sustains the properties of the original pure cluster SPT. 
That is, the mixed state $|\rho^{DW}_D\rangle\rangle$ can be regarded as the double cluster SPT, which is (strong $Z^\sigma_2$) $\otimes$  (strong $Z^\tau_2$) SPT \cite{Lee2025,Guo2024_2}.

\item {\bf Double ASPT mixed state}: For large $p_{zz}$ case corresponding to $J_{zz}>1$ in the pure state construction, 
the decohered state $|\rho^{DW}_D\rangle\rangle$ is similar to the pure state $|{\rm DW}_{\sigma SSB}\rangle$ in the sense 
that the decoherence effects become relevant (dominant). 
While the pure state $|{\rm DW}_{\sigma SSB}\rangle$ has the genuine SSB for $\sigma$-spin, the mixed state  $|\rho^{DW}_D\rangle\rangle$ is non-trivial in the sense that
the decoherence induces SWSSB. 
The emergent state possesses the strong $Z^\tau_2$ symmetry as well as weak $Z^{\sigma}_2$ symmetry, and these symmetries can be the protection symmetries of an averaged SPT
predicted in \cite{ma2023,ma2024,Ma2024_double}.
We call this state \textit{double ASPT mixed state}.
A simpler example of this phenomenon has been given in \cite{Lee2025}. 
That is, \textit{the mixed case $|\rho^{DW}_D\rangle\rangle$ is a fairly non-trivial state that has the coexisting orders of the SWSSB and ASPT}. 

\item {\bf Possibility of gapless ASPT mixed state}: 
In the pure state system for $J_{zz}=1$, gSPT, $|{\rm gSPT}\rangle$, emerges. 
We predict its mixed state counter part since in the $\sigma$-spin system before DW transformation, the $\hat{\mathcal{E}}_{\sigma^x}(p_x)$ and $\hat{\mathcal{E}}_{\sigma^z\sigma^z}$ 
induce the SWSSB phase transition as verified in \cite{Orito2025}. 
We expect that the target mixed state $|\rho^{DW}_D\rangle\rangle$ obtained by the DW duality transformation inherits its phase diagram. 
At the critical decoherence strength $p_{zz}=p^c_{zz}$ (if it exists), the decohered state $|\rho^{DW}_D\rangle\rangle$ can be regarded as a mixed-state counterpart to gSPT \cite{Guo2024_1,Guo2024_2}, namely gapless ASPT. 
Regarding the mixed state, we must be careful about what we mean by gaped and gapless for the mixed state. 
The gap that we are considering here is the energy of the least excited mode on the double state. 
The terminology ``gapless" means that there is an excitation with vanishing energy 
for $L\to \infty$ in the bulk. In this sense, this gap is different from the one recently proposed in \cite{ma2024,lessa2024,Sang2025}, 
where the gap is defined as the inverse correlation length of a quantum information theoretic correlation.
\end{enumerate}

The above predictions need to be verified. 
In the following section, we shall numerically demonstrate the emergence of the double cluster SPT as well as double ASPT phases, and
investigate how the mixed state phase transition takes place by observing the internal entanglement properties of the doubled state $|\rho^{DW}_D\rangle\rangle$.

\section{Numerical investigation by the filtering scheme with MPS}

In this section, we shall numerically study and verify the qualitative observations in the previous sections. 
To this end, we introduce the filtering scheme proposed in \cite{Orito2025} and explain the target physical observables.
We also mention a parent Hamiltonian in order to get qualitative picture of the system and then give numerical results and their physical meanings.

\subsection{Numerical filtering scheme on MPS}
We focus on the doubled system $\mathcal{H}_{u}\otimes \mathcal{H}_{\ell}$~\cite{lee2023,Lee2025}. 
The original one-dimensional $\sigma$-$\tau$ spin system is doubled and is defined on a two-leg spin-1/2 ladder. 
Then, the initial pure state $|\rho^{DW}_0\rangle\rangle$ is given by the ground state of the ladder system where each upper and lower chain Hamiltonian is given by $H^{DW}_0$, 
denoted by the ladder Hamiltonian $H^{DW}_{0,u}+H^{DW}_{0,\ell}$.  
Numerically, this ladder system can be efficiently investigated by MPS. 

On the ladder, the decoherence channel operators $\hat{\mathcal{E}}_{\tau^z\sigma^x\tau^z}(p_x)$ and $\hat{\mathcal{E}}_{\sigma^z\sigma^z}(p_{zz})$ 
act on the state $|\rho^{DW}_0\rangle\rangle$ as 
$
|\rho^{DW}_D\rangle\rangle\equiv\hat{\mathcal{E}}_{\tau^z\sigma^x\tau^z}(p_x)\hat{\mathcal{E}}_{\sigma^z\sigma^z}(p_{zz})|\rho^{DW}_0\rangle\rangle$. 
The operation represented in the exponential form can be regarded as a filtering operation \cite{Haegeman2015,Zhu2019,Orito2025}
\begin{eqnarray}
|\rho^{DW}_D\rangle\rangle= C(p_{zz},p_x,L)\prod^{L-1}_{j=0}\biggr[e^{\tau_{zz} \hat{h}^{\sigma^z\sigma^z}_{j,j+1}}e^{\tau_{x} \hat{h}^{\tau^z\sigma^x\tau^z}_{j}}\biggl]
|\rho^{DW}_0\rangle\rangle,\nonumber \\
\label{rhoD_filter}
\end{eqnarray}
where $\hat{h}^{zz}_{j,j+1}\equiv\sigma^z_{j,u}\sigma^z_{j+1,u}\otimes \sigma^z_{j,\ell}\sigma^z_{j+1,\ell}$, $\hat{h}^{\tau^z\sigma^x\tau^z}_{j}\equiv \tau^z_{j-1/2,u}\sigma^x_{j,u}\tau^z_{j+1/2,u}\otimes \tau^z_{j-1/2,\ell}\sigma^x_{j,\ell}\tau^z_{j+1/2,\ell}$ and $C(p_{zz},p_x,L)\equiv (1-2p_{zz})^{L/2}(1-2p_x)^{L/2}$. 
Note that the state $|\rho^{DW}_D\rangle\rangle$ is not a normalized state. 
Therefore, renormalization of the state vector is required in the calculation of some observables as shown later on. 
Also, as a qualitative picture, the filtering operators $\hat{h}^{zz}_{j,j+1}$ and $\hat{h}^{\tau^z\sigma^x\tau^z}_{j}$ can be regarded as additional terms 
coupling the upper and lower spins in the ladder Hamiltonian like $H^{DW}_{0,u}+H^{DW}_{0,\ell}-\sum_{j}[\lambda_{zz}\hat{h}^{zz}_{j,j+1}+\lambda_x\hat{h}^{\tau^z\sigma^x\tau^z}_{j}]$.
In fact, this filtering prescription applied to the toric code \cite{Castelnovo2008,Haegeman2015,Zhu2019,Chen2024_v2} and doubled ladder systems~\cite{Orito2025,kuno_2025,kuno_2025_v1}
has succeeded in describing physical phenomena under perturbations and decoherence. 

In the present study, we numerically perform the filtering in Eq.~(\ref{rhoD_filter}) to the MPS defined on the ladder to clarify the phase diagram of the target system. 
The decohered state $|\rho^{DW}_D\rangle\rangle$ is obtained by the (unnormalized) MPS state vector. 
Practically, this approach can be carried out efficiently by the TeNPy library \cite{TeNPy,Hauschild2024}. 
We can analyze large ladder systems and investigate the entanglement properties of the decohered state vector $|\rho^{DW}_D\rangle\rangle$. 
We also prepare an initial state $|\rho^{DW}_0\rangle\rangle$ by using the DMRG searching for the ground state of the ladder Hamiltonian $H^{DW}_{0,u}+H^{DW}_{0,\ell}$.

\subsection{Qualitative picture from a parent Hamiltonian}

Before going to physical observables, we consider a parent Hamiltonian that gives useful insight into the phase diagram of the decohered state $|\rho^{DW}_D\rangle\rangle$.
The parent Hamiltonian in its exact form has the ground state given by $|\rho^{DW}_D\rangle\rangle$.
The filtering operation in Eq.~(\ref{rhoD_filter}) provides a qualitative parent Hamiltonian picture;
the exponential form of the filtering can be regarded as effective (relevant) interactions added to the doubled Hamiltonian $H^{DW}_{0,u}+H^{DW}_{0,\ell}$, the ground state of which is given by $|\rho^{DW}_0\rangle\rangle$. 
The strengths of the effective interaction terms are proportional to $\tau_{zz}$ and $\tau_{x}$ tuned by $p_{zz}$ and 
$p_x=1/2-(1/2)(1-2p_{zz})^{1/J_{zz}}$. 
Then, the qualitative parent Hamiltonian is simply given as,
\begin{eqnarray}
H^{DW}_{D}&=&H^{DW}_{0,u}+H^{DW}_{0,\ell}\nonumber\\
&-&\lambda\sum_{j}[J_{zz}\sigma^z_{j,u}\sigma^z_{j+1,u}\sigma^z_{j,\ell}\sigma^z_{j+1,\ell}\nonumber\\
&+&\tau^z_{j-\frac{1}{2},u}\sigma^x_{j,u}\tau^z_{j+\frac{1}{2},u}\tau^z_{j-\frac{1}{2},\ell}\sigma^x_{j,\ell}\tau^z_{j+\frac{1}{2},\ell}]+\cdots , \nonumber
\end{eqnarray}
where $J_{zz}\lambda\propto\tau_{zz}$ and $\lambda\propto\tau_{x}$ \cite{Orito2025}. 
Although it is generally difficult to obtain the exact parent Hamiltonian except for the specific case \cite{Lee2025}, 
the qualitative version such as in the above often provides useful information about the filtered state~\cite{Castelnovo2008,Haegeman2015,Zhu2019,Chen2024_v2}.  
The properties of the ground state of $H^{DW}_{D}$ are close to the state $|\rho^{DW}_D\rangle\rangle$. 
Note the above qualitative Hamiltonian has the upper and lower $Z^\sigma_2\otimes Z^\tau_2$ symmetries independently, 
as it is expected for the exact parent Hamiltonian. 
For large $\lambda$, the term $\sigma^z_{j,u}\sigma^z_{j+1,u}\sigma^z_{j,\ell}\sigma^z_{j+1,\ell}$ gives a significant effect on the state $|\rho^{DW}_D\rangle\rangle$, 
and the strong $Z^\sigma_2$ symmetry tends to be spontaneously broken down to the weak one in the mixed state. 
The resultant ground state of $H^{DW}_{D}$ possesses weak $Z^\sigma_2$ as well as the strong $Z^\tau_2$ symmetries, and it
can describe an SPT phase of the original mixed state protected by the strong $Z^\tau_2$ and weak $Z^{\sigma}_2$ symmetries~\cite{Lee2025}.

\subsection{Physical observables}
To clarify the obtained state $|\rho^{DW}_D\rangle\rangle$, we introduce observables to characterize SWSSB, SPTs and entanglement properties in this section. 

The first ones are $\sigma$-spin correlators to identify the $Z^\sigma_2$ SWSSB state. 
The one of them is the partial sum of R\'{e}nyi-2 correlator, 
\begin{eqnarray}
\chi^{\rm II}&=&{2 \over L}\sum^{L/2}_{r=1}C^{\rm II}_{\sigma^z\sigma^z}(0,r),\nonumber
\label{chaiIIs}
\end{eqnarray}
where
\begin{eqnarray}
C^{\rm II}_{\sigma^z\sigma^z}(i,j)\equiv \frac{\langle\langle \rho^{DW}_D|\sigma^z_{i,u}\sigma^z_{j,u}\sigma^z_{i,\ell}\sigma^z_{j,\ell}|\rho^{DW}_D\rangle\rangle}{\langle\langle \rho^{DW}_D|\rho^{DW}_D\rangle\rangle}.\nonumber
\end{eqnarray}
Through the Choi map, it is not so difficult to show that the correlator $C^{\rm II}_{\sigma^z\sigma^z}(i,j)$ is nothing but  
$\displaystyle{C^{\rm II}_{\sigma^z_i \sigma^z_j}= [\Tr[\sigma^z_i\sigma^z_j\rho^{DW}_D \sigma^z_j\sigma^z_i \rho^{DW}_D]]/ [\Tr[(\rho^{DW}_D)^2]]}$. 
This observable is an order parameter that detects SSB of the strong $Z^{\sigma}_2$ symmetry but {\it not} the weak symmetry~\cite{lee2023,lessa2024,sala2024,Ma2024_double}.
In fact, $C^{\rm II}_{\sigma^z\sigma^z}(i,j)\neq 0$ for $|i-j| \to \infty$ indicates the emergence of a strong-symmetry-breaking state but not a genuine SSB. 

Another observable is the partial sum of the $\sigma^z\sigma^z$-correlator of the upper chain,
\begin{eqnarray}
\chi^{\rm I}=\frac{2}{L}\sum^{L/2}_{r=1}C^{\rm I}_{\sigma^z,st}(0,r),
\end{eqnarray}
where 
$
C^{\rm I}_{\sigma^z,st}(i,j)
$ is a correlator searching the $Z^{\sigma}_2$ genuine SSB in the doubled Hilbert space and given by,
$$
C^{\rm I}_{\sigma^z,st}(i,j)=\frac{\langle\langle {\bf 1}|\sigma^z_{i,u}\sigma^z_{j,u}|\rho^{DW}_D\rangle\rangle}{\langle\langle {\bf 1}|\rho^{DW}_D\rangle\rangle},
$$ 
where $|{\bf 1}\rangle\rangle$ is the relative state \cite{Ma2024_double} defined by $|{\bf 1}\rangle\rangle\equiv \displaystyle{\frac{1}{2^{3L/2}}\prod^{L-1}_{j=0}|t\rangle_j}$ with $|t\rangle_j=|\uparrow_u\uparrow_{\ell}\rangle_j+|\downarrow_u\downarrow_{\ell}\rangle_j$, and the corresponding quantity in the original physical Hilbert space is 
${\rm Tr}[\rho^{DW}_D \sigma^z_{i}\sigma^z_{j}]$. 
This quantity $\chi^{\rm I}$ is used as an order parameter of the genuine SSB~\cite{lessa2024}. 
If the genuine SSB occurs by some additional effect, the mixed state $|\rho^{DW}_D\rangle\rangle$ is no longer $Z^\sigma_2$ symmetric even in the weak sense.
In the doubled Hilbert space, the $Z^\sigma_2$ SWSSB is characterized by  $\chi^{\rm II}\sim \mathcal{O}(1)$ and 
$\chi^{\rm I}\sim 0$. 

As an order parameter of SPT order, we introduce a non-local R\'{e}nyi-2 string correlator playing an important role for identifying ASPT state~\cite{Lee2025}, 
\begin{eqnarray}
S^{(2)}_t\equiv \frac{\langle \langle \rho^{DW}_{D}|\tau^z_{\frac{1}{2},u}\tau^z_{\frac{1}{2},\ell}\biggl[\prod^{k}_{j=1}\sigma^x_{j,u}\sigma^x_{j,\ell}\biggr]\tau^z_{k+\frac{1}{2},u}\tau^z_{k+\frac{1}{2},\ell}|\rho^{DW}_D\rangle\rangle}{\langle \langle \rho^{DW}_D|\rho^{DW}_D\rangle\rangle},\nonumber\\
\end{eqnarray}
the counterpart of which in the original mixed state representation is 
essentially given as
$
\mbox{Tr} \Big[\rho^{DW}_D \tau^z_{{1 \over 2}}\Big(\prod^k_{j=1}\sigma^x_j\Big) \tau^z_{k+{1\over2}}\rho^{DW}_D
\tau^z_{{1 \over 2}}(\prod^k_{j=1}\sigma^x_j) \tau^z_{k+{1\over2}}\Big].
$

We also measure the ordinary string correlator,
\begin{eqnarray}
S^{(1)}_t\equiv \frac{\langle \langle {\bf 1}|\tau^z_{\frac{1}{2},u}\biggl[\prod^{k}_{j=1}\sigma^x_{j,u}\biggr]\tau^z_{k+\frac{1}{2},u}|\rho^{DW}_D\rangle\rangle}{\langle \langle {\bf 1}|\rho^{DW}_D\rangle\rangle},
\end{eqnarray}
the counterpart of which in the original mixed state representation is given as 
$
\mbox{Tr} \Big[\rho^{DW}_D\tau^z_{{1 \over 2}}\Big(\prod^k_{j=1}\sigma^x_j\Big) \tau^z_{k+{1\over2}}\Big]
$.
The combination of $S^{(2)}_{t}$ and $S^{(1)}_{t}$ can characterize the bulk ASPT in the state $|\rho^{DW}_D\rangle\rangle$. 
That is, $S^{(2)}_{t}\neq 0$ for $|i-j| \to \infty$ and $S^{(1)}_{t}= 0$ indicates the emergence of the double ASPT phase protected by 
the strong $Z^\tau_2$ symmetry and weak $Z^{\sigma}_2$ symmetry \cite{Lee2025}.   

\begin{figure}[t]
\begin{center} 
\includegraphics[width=7cm]{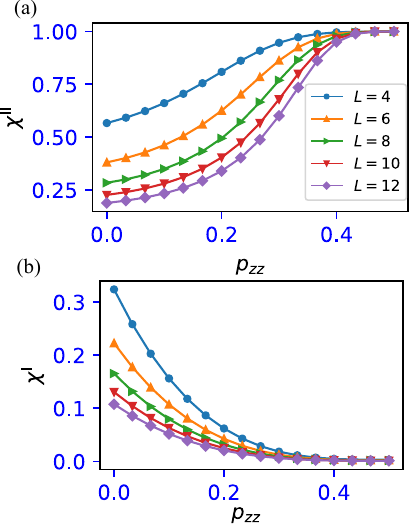}  
\end{center} 
\caption{Behaviors of $\chi^{\rm II}$ [(a)] and $\chi^{\rm I}$[(b)] for various system sizes.}
\label{Fig_corr}
\end{figure}
As third observable, we introduce entanglement entropy (EE) for a subsystem (subsystem A) for the state $|\rho^{DW}_D\rangle\rangle$, 
$$
S_A=-\Tr_{A}[\tilde{\rho}_{D,A}\ln \tilde{\rho}_{D,A}],
$$
where $\tilde{\rho}_{D,A}=\Tr_{\bar{A}}\tilde{\rho}_D$ with subsystem $\bar{A}$ complement to $A$, and 
$\tilde{\rho}^{DW}_D=|\tilde{\rho}^{DW}_D\rangle\rangle \langle\langle \tilde{\rho}^{DW}_D|$ 
and $|\tilde{\rho}^{DW}_D\rangle\rangle$ is the normalized state of $|\rho^{DW}_D\rangle\rangle$, $|\tilde{\rho}^{DW}_D\rangle\rangle
=|\rho^{DW}_D\rangle\rangle/\sqrt{\langle\langle \rho^{DW}_D|\rho^{DW}_D\rangle\rangle}$.
This quantity measures the entanglement of the doubled state similarly to the usual pure-state case.
We also observe the low-lying entanglement spectrum (ES) of $S_A$.

\subsection{Numerical results}
In this section, we display the numerical results. 
In numerical calculations, we prepare the initial state $|\rho^{DW}_0\rangle\rangle$ with $J_{zz}=0.6$, which is a pure double cluster SPT state produced by the DMRG.
In all numerical simulations, we set maximum bond dimension $D=100$-$200$ and truncate the singular value less than $\mathcal{O}(10^{-6})$, and the energy convergence of the iterative DMRG sweeping is $\Delta E < \mathcal{O}(10^{-4})$ to obtain the initial MPS ground state.
The filtering operation to the state $|\rho^{DW}_0\rangle\rangle$ is efficiently carried out by the TeNPy package \cite{Hauschild2024}. 
We then obtain the state $|\rho^{DW}_D\rangle\rangle$ as varying $p_{zz}$ and $p_x$ is determined by $p_x=1/2-(1/2)(1-2p_{zz})^{1/J_{zz}}$ as mentioned in the previous section. 

Figures \ref{Fig_corr} (a) and \ref{Fig_corr} (b) show results of $\chi^{\rm II}$ and $\chi^{\rm I}$. 
For small $p_{zz}$, $\chi^{\rm II}$ is small but has a finite value and $\chi^{\rm I}$ has a fairly large value due to the remaining initial effects of a finite term $J_{zz}$ in the initial state. 
As $p_{zz}$ increases, $\chi^{\rm II}$ increases while $\chi^{\rm I}$ decreases and approaches zero. 
This behavior is a signal of the $Z^\sigma_{2}$ SWSSB.

\begin{figure}[t]
\begin{center} 
\includegraphics[width=7cm]{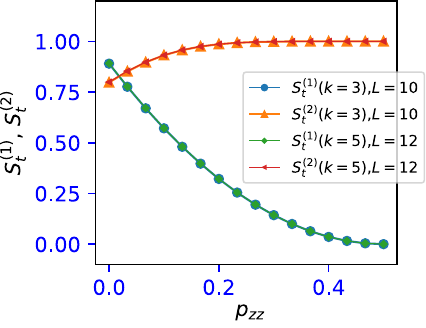}  
\end{center} 
\caption{Behaviors of $S^{(1)}_t$ and $S^{(2)}_t$.
We set $L=10$ and $12$.}
\label{Fig_STO}
\end{figure}

Next, we show the results of the string order parameters $S^{(1)}_t$ and $S^{(2)}_t$. 
The results for $L=10$ for cases $k=4$ and $L=12$ with the cases $k=5$ are shown in Fig.~\ref{Fig_STO}. 
The conventional string order parameter $S^{(1)}_t$ decays as $p_{zz}$ increases since the decoherence $\hat{\mathcal{E}}_{\sigma^z\sigma^z}$ 
erases the influence of the $\tau^z\sigma^x\tau^z$ term in the initial doubled state and reduces the strong ASPT phase. 
On the other hand, the R\'{e}nyi-2 string correlator $S^{(2)}_t$ continues to be finite enough for any $p_{zz}$ and approaches to one as $p_{zz}\to 1/2$. 
Thus, for large $p_{zz}$, the decohered state is the strong $Z^\tau_2$ and weak $Z^{\sigma}_2$ SPT.

\begin{figure}[t]
\begin{center} 
\includegraphics[width=8cm]{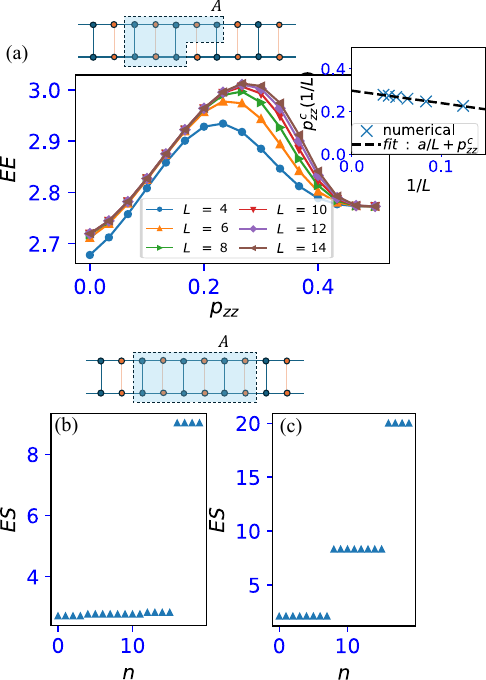}  
\end{center} 
\caption{(a) EE for a fixed subsystem. 
Upper panel: Entanglement cut independent to system size. Right panel: Extraction of the transition point. To estimate locations of peaks of the EE, we employed a sixth-order polynomial function to estimate $p_{zz}^c$ for each system size. Using the data of finite size systems and a first-order polynomial function, $a/L+p^c_{zz}$, where $a$ and $p^c_{zz}$ are fitting parameters, we estimated the phase transition point in the thermodynamic limit. (b) and (c): The low-lying $n$-th ES for $p_{zz}=0.05$ [(b)] and $p_{zz}=0.45$ [(c)]. The entanglement cuts are shown in the upper panel. The subsystem $A$ includes the upper and lower $\sigma$ and $\tau$-spins from $j=0$ to $j=L/2-1$ in the system. The four-bonds are cut by the boundary of the subsystem $A$. We set $L=10$.}
\label{Fig_EE}
\end{figure}

We further study the entanglement properties of the doubled system. 
As the first approach, we set a subsystem $A$ that includes eight connected $\sigma$-$\tau$ spins as shown in the inset of Fig.~\ref{Fig_EE}. 
The behavior of EE as a function of $p_{zz}$ is shown in the upper of Fig.~\ref{Fig_EE} (a). 
EE exhibit peaks, the heights are different for different system sizes, and the dependence seems to exhibit sub-volume law, implying a criticality. 
These observations indicate that the entanglement phase transition occurs corresponding to the change from the strong ASPT to the double ASPT. 
From the data, we estimate the transition strength $p^c_{zz}$ as $p^c_{zz}=0.297$. 
Additionally, we investigate the subsystem-size dependence of the EE for the doubled system at criticality as shown in Appendix B.

We further observe the low-lying ES from the EE and set the subsystem $A$ to the half-subsystem shown in the upper of Fig.~\ref{Fig_EE}(b). 
In this setup, the ES has a contribution from the cut of four bonds at the boundary of the subsystem $A$.
For typical values of $p_{zz}$, we plot the ES. 
At $p_{zz}=0.05$, where the state is in the deep strong ASPT phase, the ES exhibits nearly 16-fold degeneracy as seen in Fig.~\ref{Fig_EE} (b). 
This indicates that the decohered state preserves the edge-mode degeneracy of the pure double cluster SPT states. 
Combined with the numerical results of the bulk string correlations, we conclude that the bulk-edge correspondence holds in the doubled state.

We turn to the observation of the data for $p_{zz}=0.45$, where the state is in the deep double ASPT phase. 
As shown in Fig.~\ref{Fig_EE} (c), the ES exhibits an eight-fold degeneracy. 
This is a signature that for this doubled decohered state in the ladder system,  
the edge mode degeneracy can be eight, that is, eight independent edge modes protected by the strong $Z^\tau_2$ symmetry and the weak $Z^{\sigma}_2$ symmetry
(denoted by $Z^{\tau}_{2,u}\otimes Z^{\tau}_{2,\ell} \otimes Z^{\sigma}_{2}(\mbox{weak})$ symmetry) can exist if we set open boundaries. This should be compared with the above $p_{zz}=0.05$ case, in which 
$Z^{\tau}_{2,u}\otimes Z^{\tau}_{2,\ell} \otimes Z^{\sigma}_{2,u}\otimes Z^{\sigma}_{2,l}$ symmetry exists and the form of the edge Hamiltonian
is restricted more strictly producing sixteen edge modes as a result. 
Such a symmetry-protected edge mode has been predicted in a general theoretical level \cite{Ma2024_double}.

\section{Detailed comments on co-existance of ASPT and SWSSB}

In the previous section, we showed that the double ASPT state emerges as the strength of decoherence increases beyond the phase transition point.
In that phase, ASPT and SWSSB coexist as observed by the R\'{e}nyi-2 correlators, R\'{e}nyi-2 string order and EE.
Here, we discuss the relationship between this observation and previous works that studied the ASPT and SWSSB \cite{ma2024,lessa2024}
as the above result may seem odd. 
There are two concerns.

The first one concerns classification of ASPT.
In Ref. \cite{ma2024}, from the point of view of the pure state SPT state, the ASPT state is classified by symmetrically invertible property.
For the present model, we expect that states under weak decoherence inherit this property as the initial state is the genuine SPT cluster state.
As the result of Fig.~\ref{Fig_corr} shows, the R\'{e}nyi-2 correlator behaves as $C^{\rm II}_{\sigma^z\sigma^z} (i,j) \to 0$ for the infrared (IR) limit $L, |i-j| \to \infty$.
Provided that the fidelity correlator \cite{lessa2024} and R\'{e}nyi-2 correlator exhibit qualitatively the same behavior in the IR limit, SWSSB does not
occur there as discussed in ~\cite{lessa2024}, and this weakly decohered phase is an ASPT state of the above classification \cite{ma2024}.
However, as the strength of decoherence increases, a phase transition occurs and the double ASPT state appears.
As a result of this transition, the resultant state may lose the symmetrically invertible property.
In fact, the numerical result $C^{\rm II}_{\sigma^z\sigma^z} (i,j) \to c\neq 0$ in Fig.~\ref{Fig_corr} indicates the appearance of SWSSB.
As Ref.~\cite{lessa2024} showed, mixed states with SWSSB is not symmetrically invertible. 
From this observation, we think that the double ASPT state exhibits SWSSB and is not symmetrically invertible, indicating that
the double SPT state is not to be classified by the scheme of Ref.~\cite{ma2024} (a similar observation appeared in \cite{Guo2024_2}).
None the less, we call the strongly-decohered state double ASPT state, as it exhibits the R\'{e}nyi-2 string order as well as EE, which indicates
non-trivial edge modes.
Detailed investigation and deep understanding of this interesting phenomenon are future problem.

Second one concerns ``gap" of the system. 
For a density matrix, it is generally difficult to define the gap. 
However as showed by Ref.~\cite{ma2024}, the ASPT can have a gap in the sense of the inverse of the correlation length of a quantum information theoretic correlation \cite{Sang2025}, 
while a SWSSB has no gap in this sense shown in Ref.~\cite{lessa2024}. 
This discrepancy should be clarified. 
Here, we should comment again on the meaning of gap in this work, that is, we focus on the doubled purified (un-normalized) state defined on the doubled Hilbert space. 
Gaped state in this formalism is the one that any excitation on the doubled decohered state (including the multiplet coming from the SWSSB) has a finite energy gap for $L\to \infty$. 
Thus, this notion of gap does not necessarily equate to those in \cite{lessa2024,Sang2025}.
One interesting question is whether the double ASPT numerically observed here is an exceptional example as the ASPT proposed in \cite{ma2023,ma2024} or not. 
This theoretical question can be future work. 

\section{Summary and conclusion} 
In this work, we proposed a construction scheme for strong symmetry-protected mixed SPT  and double ASPT states under decoherence channels, 
based on the pure gSPT framework and the DW duality transformation. 
We predicted that the double ASPT phase exhibits coexisting orders--an SWSSB and an ASPT—protected by one strong symmetry and one weak symmetry. 
Furthermore, following \cite{Orito2025}, we predicted a mixed-state phase transition between the strong ASPT and the double ASPT phases as the strength of the quantum channel varies.

To validate these predictions, we performed numerical simulations using the filtering scheme applied to MPSs. 
We characterize the strong ASPT and double ASPT phases through correlation functions and investigate their phase transition by analyzing the entanglement 
properties of the doubled decohered state on a ladder system. 
Additionally, we showed that the entanglement spectrum reveals edge mode degeneracy, distinguishing the strong ASPT from the double ASPT phase, in agreement with previous 
theoretical predictions \cite{Guo2024_1,Guo2024_2}.

We also point out that more simplified quantum channels can exhibit strong ASPT and double ASPT phases, as discussed in Appendix C. 

Our findings imply that multiple decoherence processes can give rise to a rich variety of nontrivial and intrinsic mixed states, particularly, ASPT phases. 
More broadly, our quantum-channel-based approach could be extended to pure topologically ordered states, such as the toric code, by designing suitable decoherence channels. 
The recent extensions of 1-form symmetry to weak or averaged symmetry notions in the toric code \cite{zhang2024,sohal2024,kuno_2025_v1,Wang_2025} 
suggest promising directions for future research.

In addition, we focus on the partial sum of the R\'{e}nyi-2 correlator, $\chi^{\rm II}$. 
This quantity can be related to the the fidelity susceptibility \cite{Zhang2024_FDT}. 
Recently, the fidelity susceptibility is further  related to the fluctuation-dissipation theorem \cite{Zhang2024_FDT}. 
It is an interesting problem to investigate the relationship between $\chi^{\rm II}$ and the fluctuation-dissipation theorem in the mixed phases that we found in this work.

\section*{Acknowledgements}
This work is supported by JSPS KAKENHI: JP23K13026(Y.K.) and JP23KJ0360(T.O.). 

\section*{Data availability}
The data and code that support the main findings of this study will be available on Zenodo. 

\renewcommand{\thesection}{A\arabic{section}} 
\renewcommand{\theequation}{A\arabic{equation}}
\renewcommand{\thefigure}{A\arabic{figure}}
\setcounter{equation}{0}
\setcounter{figure}{0}

\appendix
\section*{Appendix}
\section{Strong and weak $Z_2$ symmetries}
In general, two types of symmetries can be introduced in mixed states, namely, strong and weak symmetries \cite{Buca2012,groot2022}.
In this work, we consider its $Z_2$ symmetry version. Suppose the generator of which is $\{\hat{1},U_{Z_2}\}$ with $U^2_{Z_2}=\hat{1}$, 
the strong symmetry for a density matrix is defined as $U_{Z_2}\rho=e^{i\theta}\rho$, where $\rho$ is a symmetric mixed state and $\theta$ is a global phase factor, $\theta\in \{0,\pi\}$. 

We further introduce the weak-symmetry condition defined as  
$
U_{Z_2}\rho U^\dagger_{Z_2} = \rho
$. 
This symmetry is satisfied on the average level \cite{ma2024}, in the sense that the symmetry is satisfied after taking the ensemble average in the density matrix.

Strong and weak symmetry conditions are further defined for quantum channel. 
The Kraus operator form of channel is given as \cite{Nielsen2011}
$\mathcal{E}(\rho)=\sum^{N-1}_{\ell=0}K_{\ell} \rho K^\dagger_{\ell}
$, where $\{K_{\ell}\}$ are a set of Kraus operators satisfying $\sum^{N-1}_{\ell=0} K^\dagger_{\ell} K_{\ell}=\hat{I}$ with $\hat{I}$ being the identity operation. 
The quantum channel $\mathcal{E}$ induces changes in mixed states. 
Here, the strong $Z_2$-symmetry condition on the channel is given as 
$K_{\ell}U_{Z_2}=e^{i\theta} U_{Z_2} K_{\ell}$ for any $\ell$. 

On the other hand, the weak symmetry condition on the channel is expressed as 
$
U_{Z_2}\biggl[\sum_{\ell}K_{\ell} \rho K^\dagger_{\ell}\biggr]U^\dagger_{Z_2}=\mathcal{E}(\rho)$.
This condition does not require that each Kraus operator commutes with non-trivial generator $U_{Z_2}$. 
A channel that satisfies the strong-symmetry condition is automatically weak symmetric.

\section{Subsystem-size dependence of entanglement entropy in the vicinity transition point.}

\begin{figure}[t]
\begin{center} 
\vspace{0.5cm}
\includegraphics[width=7cm]{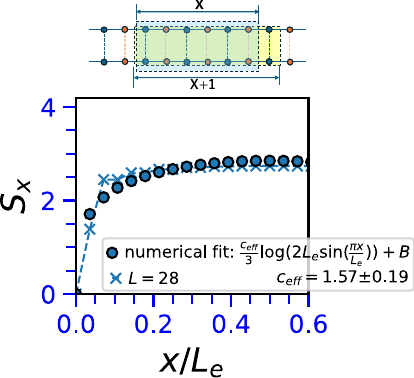}  
\end{center} 
\caption{Scaling of $S_x$  for $p_{zz}=0.29$.
$x$ is the number of rung in subsystem, where its total number is $L_e=28$ ($L=14$). 
}
\label{FigA3}
\end{figure}
It is widely recognized that the scaling of EE in on-critical states obeys the logarithmic scaling \cite{CC2004}. 
In order to very this, we show how the scaling of EE at the critical point, estimated in the main text, is affected by decoherence.
To examine whether the scaling of EE obeys the logarithmic scaling with respect to the subsystem size, 
we employ the following fitting function ~\cite{CC2004}, 
\begin{eqnarray}
S_x=\frac{c_{\rm eff}}{3}\log(2L_e\sin(\pi x/L_e))+B,
\label{CC}
\end{eqnarray}
where $c_{\rm eff}$ and $B$ are fitting parameters, $x$ is the length of the subsystem and $L_e$ is the total number of rung in the doubled system (see the upper schematic of Fig.~\ref{FigA3}).
Here, $c_{\rm eff}$ corresponds to the effective central charge.  
That is, we use the two vertical entanglement cuts in the periodic system. 
Technically speaking, Eq.~(\ref{CC}) is employed to estimate entanglement scaling for one-dimensional systems.

Figure~\ref{FigA3} shows the result $S_x$ for the system at $p_{zz}=0.29$. 
The data points of $S_x$ for the quantum state exhibit a form approximately similar to that of one-dimensional chain system, 
represented by the fitting function of Eq.~(\ref{CC}), although the estimated central charge $c_{\rm eff}$ deviates from the one of Ising CFT ($=1$). 
Thus, this data supports that the transition point we found in the main text exhibits near logarithmic growth of the EE.



\section{Simplified case for emergence of double ASPT}
We observe the the strong ASPT and double ASPT in more simplified decoherence channel although these appearances have less theoretical backgrounds 
for the presence of the $Z^{\sigma}_2$ SWSSB followed by the previous study \cite{Orito2025}. 
But, we numerically observe them. 

Indeed, we only consider the channel $\hat{\mathcal{E}}_{\sigma^z\sigma^z}(p_{zz})$ and its decohered state given by 
$|\rho^{DW'}_0\rangle\rangle\equiv \hat{\mathcal{E}}_{\sigma^z\sigma^z}(p_{zz})|\rho^{DW}_0\rangle\rangle$. 
We numerically confirmed the emergence of the strong ASPT and double ASPT phases by observing the same quantities considered in the main text. 
Thus, the condition between $p_{zz}$ and $p_x$ can be relaxed and various settings of quantum channels can exhibit similar mixed phases shown in the main text.

\bibliography{ref}

\end{document}